\documentclass[%
 reprint,
 superscriptaddress,
 amsmath,amssymb,
 aps,longbibliography
]{revtex4-1}

\usepackage{graphicx}
\usepackage{dcolumn}
\usepackage{bm}
\usepackage{hyperref}
\usepackage[mathlines]{lineno}
\usepackage[bottom]{footmisc}
\usepackage[super]{nth}%
\usepackage{color}
\usepackage{hhline}

\begin{document}
\raggedbottom

\title{Optical rogue waves in multifractal photonic arrays}
\author{F. Sgrignuoli}
\affiliation{Department of Electrical and Computer Engineering, Boston University, 8 Saint Mary's Street, Boston, Massachusetts 02215, USA}
\author{Y. Chen}
\affiliation{Department of Electrical and Computer Engineering, Boston University, 8 Saint Mary's Street, Boston, Massachusetts 02215, USA}
\author{S. Gorsky}
\affiliation{Department of Electrical and Computer Engineering, Boston University, 8 Saint Mary's Street, Boston, Massachusetts 02215, USA}
\author{W. A. Britton}
\affiliation{Division of Materials Science \& Engineering, Boston University, 15 Saint Mary's St. Brookline, Massachusetts, 02446, USA}
\author{L. Dal Negro}
\email{dalnegro@bu.edu}
\affiliation{Department of Electrical and Computer Engineering, Boston University, 8 Saint Mary's Street, Boston, Massachusetts 02215, USA}
\affiliation{Division of Material Science and Engineering, Boston University, 15 Saint Mary's Street, Brookline, Massachusetts 02446, USA}
\affiliation{Department of Physics, Boston University, 590 Commonwealth Avenue, Boston, Massachusetts 02215, USA}
\begin{abstract}
Optical rogue waves are demonstrated in the far-field scattered radiation from photonic arrays designed according to the aperiodic distributions of prime elements in complex quadratic fields. Specifically, by studying light diffraction from Eisenstein and Gaussian prime arrays we establish a connection between the formation of optical rogue waves and multifractality in the visible single-scattering regime. We link strong multifractality with the heavy-tail probability distributions that describe the fluctuations of scattered radiation from the fabricated arrays. Our findings pave the way to control high-intensity rogue waves using deterministic arrays of dielectric nanostructures for enhanced sensing and lithographic applications. 
\end{abstract}

\pacs{Valid PACS appear here}
\keywords{Suggested keywords}
\maketitle
The term rogue waves (RWs) was originally introduced in hydrodynamics to describe the behavior of giant waves that emerge unexpectedly on a relative clam ocean releasing exceptionally destructive power \cite{Kharif_2}. Since then, RWs were observed in different contexts (e.g., optics \cite{Dudley,Dudley_2,Solli,Arecchi,Hohmann,Mathis,Safari}, condensed matter physics \cite{Bludov}, optical turbulence \cite{Walczak}, and even in finance \cite{Zhen}, to cite a few) becoming an important subject of interdisciplinary research \cite{Dudley,Dudley_2,Ruban}. 

Despite the theoretical and experimental efforts of the last twelve years \cite{Dudley,Dudley_2,Dysthe}, an exact definition of RW does not exist yet \cite{Ruban}. Moreover, as pointed out in recent reviews \cite{Dudley,Dudley_2}, the analogy between optical and ocean rogue waves must be handled with care. However, a common feature of all these studies is the presence of heavy-tailed probability density functions (PDFs) describing the intensity fluctuations of the waves. Therefore, RWs are characterized by non-Gaussian statistics, implying that waves with extremely large amplitudes appear more often than what predicted from the normal distribution. Heavy-tailed statistical distributions arise in the context of extreme value theory (EVT) beyond the validity of the central limit theorem when limit-processes, such as the sum of dependent and correlated variables, are considered \cite{Coles,Metzger}. Fundamental aspects of the physics of optical RW are the presence of heavy-tails PDFs in the statistical distribution of scattered radiation and its connection with the structural properties of complex systems. However, the mechanisms driving the formation of RWs is a matter of debate and depends on the particular system under study \cite{Kharif,Kharif_2,Dudley_2,Onorato}. 

Recently, Dematteis et. al. proposed and tested a statistical theory demonstrating that water tank rogue waves are hydrodynamic instantons \cite{Dematteis}. Instead, optical RWs were demonstrated for the first time by Solli et. al. in microstructured optical fibers driven within a noise-sensitive nonlinear regime \cite{Solli}. The strong phase gradients and fluctuations that give rise to natural focusing phenomena \cite{Sharma}, described by the theory of catastrophe optics \cite{Berry,Nye}, also produce abrupt, rare, and extreme fluctuations of field amplitudes. These studies established a connection between rogue-type behavior and far-field diffracted caustics \cite{Mathis,Safari,BerryReview}. Moreover, these discoveries showed that nonlinearity is not essential for the generation of optical RWs that can be observed in linear systems when a suitable random phase structure is imparted on a coherent optical field \cite{Mathis,Safari,Hohmann} or when disordered phases exhibit long-range correlations in space \cite{BonattoSR}. Wave focusing due to collective effects in a correlated complex medium \cite{Metzger}, granularity, and spatial inhomogeneity \cite{Arecchi,Onorato} have been identified as the main factors for the occurrence of RWs in the optical regime. 

In this work, we propose and demonstrate a novel approach for the generation of optical RWs based on the engineering of deterministic arrays of dielectric nanostructures with aperiodic multifractal geometry \cite{Parisi}. Multifractals, i.e., intertwined sets of self-similar structures, are inhomogeneous systems characterized by complex fluctuations over multiple-length scales that encode long-range correlations \cite{Sharma}. Introduced by Frisch and Parisi to analyze the multi-scale energy dissipation in turbulent fluids \cite{Parisi}, multifractality (MF) became an interdisciplinary concept that is investigated in various fields of research. Besides finance \cite{Schmitt}, chaotic systems \cite{Paladin}, and condensate matter physics \cite{Nakayama}, self-similarity and multifractality have been observed in soliton-based systems \cite{Soljacic}--like the ones used to demonstrate optical RWs (see, e.g., \cite{Dudley} and references therein)-- and in extremely rare natural hazards \cite{Sharma}, such as tsunami \cite{Telesca}, earthquakes \cite{Hirabayashi}, and oceanographic rogue waves \cite{Hadjihosseini}. Interestingly, also diffracted caustics, as pointed out by Berry and Upstill \cite{BerryReview}, are characterized by a hierarchy of self-similar length-scales.

Motivated by these findings, we ask whether a fundamental connection exists between MF and RWs in linear optics. To establish such a relation, we measured the diffraction intensity patterns produced by aperiodic photonic arrays designed from fundamental structures of algebraic number theory \cite{Sgrignuoli_MF,Wang_Prime,Lang,Dekker} that inherit the multifractality of the distributions of prime elements \cite{Wolf}. The multifractality of these photonic systems has been demonstrated in ref.\,\cite{Sgrignuoli_MF} in the multiple scattering regime by performing leaky-mode imaging experiments at high numerical aperture. Instead, in the present work we systematically investigate their single scattering properties focusing on the far-field diffracted radiation from Eisenstein and Gaussian prime arrays. First of all, we demonstrate that the structure factor of these arrays, which is proportional to the far-field scattered intensity \cite{Goodman1}, exhibits strong multifractal behavior. As a second step, we show that the PDFs of scattered radiation, which characterize the spatial fluctuations of coherent laser light diffracted by the aperiodic structures, as well as the distributions of the most intense values are described by Pareto-type and Fr\'echet-type extreme value distributions, respectively. Finally, we determine that these non-Gaussian statistics originate from the strong multifractal geometry of the investigated photonic arrays.

\begin{figure}[t!]
\centering
\includegraphics[width=\columnwidth]{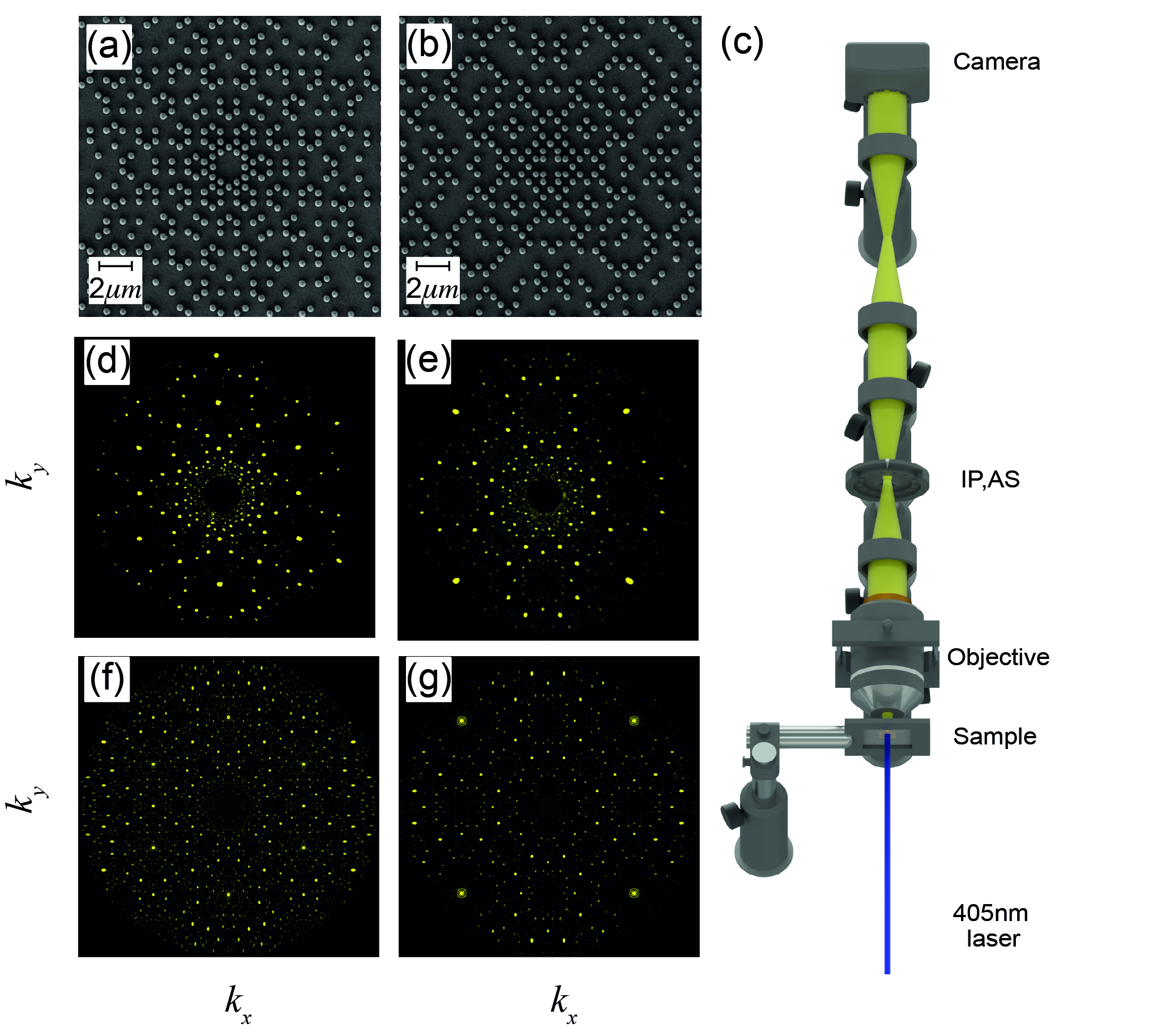}
\caption{SEM images of the fabricated photonic arrays arranged in an Eisenstein (a) and Gaussian (b) geometry, respectively. Nanocylinders have a 210\,nm mean diameter, 250\,nm height, and average inter-particle separation of 650\,nm. Note that unavoidable fabrication imperfections, such as surface roughness and fluctuations in the size and position of the particles (see Supplemental Material for more information), are deeply sub-wavelength and do not significantly influence the diffraction patterns measured using the customized optical setup shown in panel (c). IP image plane, AS Aperture stop. Measured (d-e) and calculated (f-g) intensity profiles of the prime arrays reported at the top of each column. The diffraction pattern (f) and (g) are calculated by considering more than 10$^4$ prime elements.}
\label{Fig1}
\end{figure}
 
The photonic structures are  fabricated using electron beam lithography. Specifically, TiO$_2$ nanocylinders deposited atop a transparent SiO$_2$ substrate are arranged as the prime elements of the Eisenstein and Gaussian integers. Recently, these photonic arrays were introduced to exploit structural multifractality as an engineering approach for optical sensing, lasing, and multispectral devices \cite{Wang_Prime,Sgrignuoli_MF}. Scanning electron microscope (SEM) images of the fabricated devices are reported in Figs.\,\ref{Fig1}\,(a) and (b). More details on the fabrication as well as on their geometrical properties are discussed in ref.\,\cite{Sgrignuoli_MF} and in the Supplemental Material. The experimental setup used to measure the diffraction pattern of laser light scattered by the arrays is shown in Fig.\,\ref{Fig1}\,(c). A 405\,nm laser is focused onto the device to uniformly illuminate the sample. The forward scattered light is collected by a high numerical aperture objective (NA=0.9 Olympus MPlanFL N) that gathers light scattered up to $64^\circ$ from the normal direction. Immediately behind the objective, a 4-F optical system creates an intermediate image plane and an intermediate Fourier plane. An iris, located at the intermediate image plane, was used to restrict the light collection area only to the patterned regions. The intermediate Fourier plane was re-imaged onto a CCD with the appropriate magnification by using a second 4-F optical system. Finally, digital filtering was employed to remove the strong direct component of the diffraction spectra to produce the clear images reported in Figs.\,\ref{Fig1}\,(d-e) for the Eisenstein and Gaussian configuration, respectively.  
The experimental results are compared with the predicted far-field diffraction intensity that is proportional to the computed structure factor of the investigated arrays, defined as \cite{Goodman1,Sgrignuoli_MF}:
\begin{equation}
S(\bm{k})=\frac{1}{N}\Biggl|\sum_{j=1}^N e^{-i\bm{k}\cdot\bm{r}_j}\Biggr|^2
\end{equation} 
where $\bm{k}$ is the in-plane component of the wavevector and $\bm{r}_j$ are the vector positions of the $N$ nanoparticles in the array. The computed structure factors are displayed in Fig.\,\ref{Fig1}\,(f-g). We found a good agreement between the computed and the experimentally measured diffraction spectra, whose peaks are slightly broadened due to unavoidable fabrication imperfections. We have quantified these effects in the Supplemental Material where we also provide a detailed analysis based on the comparison of the pair distribution functions of the arrays.

\begin{figure}[b!]
\centering
\includegraphics[width=\columnwidth]{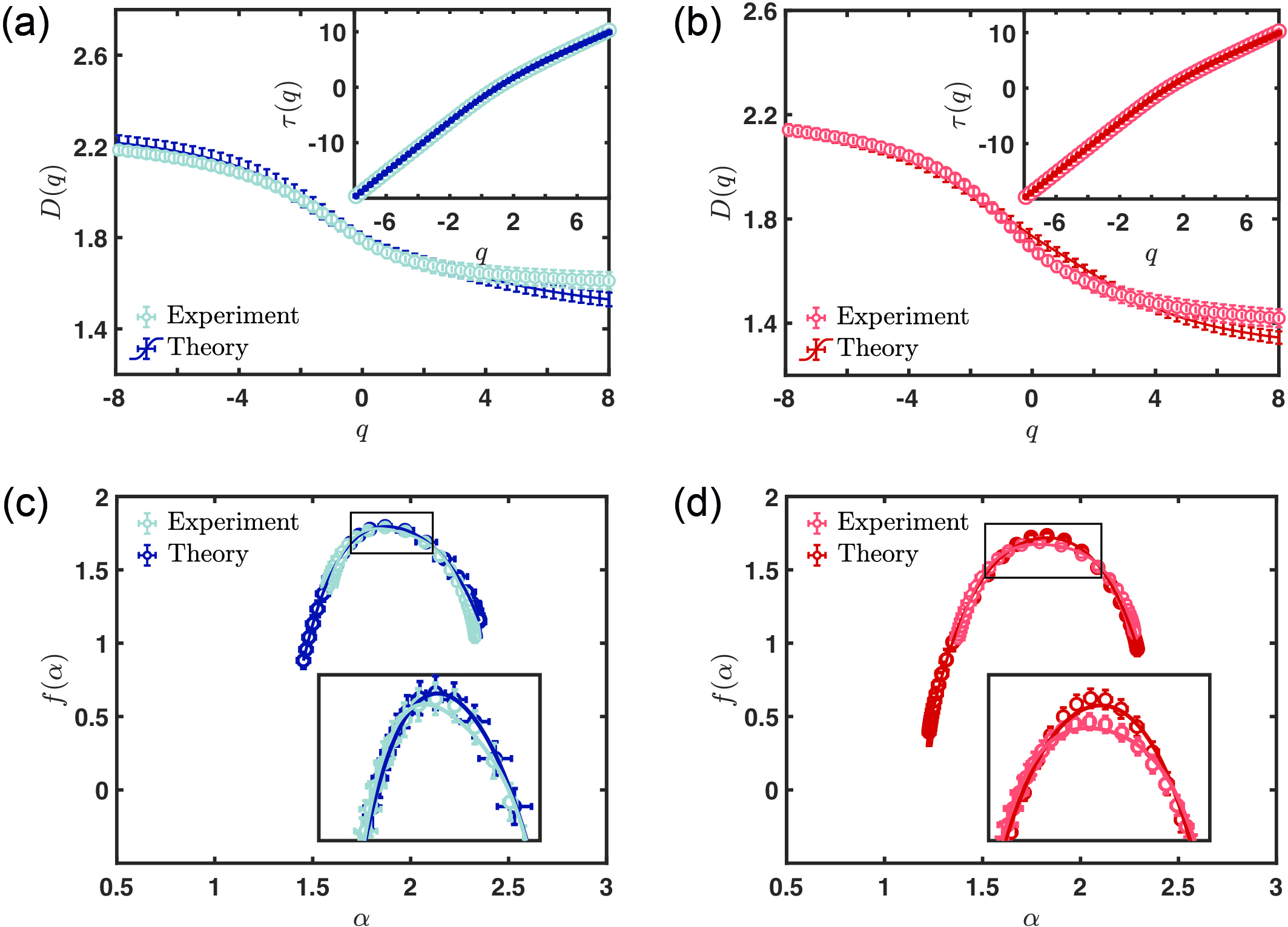}
\caption{$D(q)$ as a function of the moments $q$ extrapolated from the measured (pastel-blue and pastel-red) and simulated navy-blue and dark-red) scattered radiation from the Eisenstein (a) and Gaussian (b) arrays, respectively. The insets report the $\tau(q)$ exponents. Panels (c) and (d) show the $f(\alpha)$ spectra of the diffracted intensity of  the Eisenstein and Gaussian array, respectively. Markers represent the data, while the continuous lines refer to the best fits obtained by using a square-least method based on the polynomial function $f(\alpha)=p_1+p_2\alpha+p_3\alpha^2+p_4\alpha^3+p_5\alpha^4$. Error bars take into account the different threshold percentages (between 55\% and 75\% of the maximum intensity value) used to binarize the diffraction patterns reported in Fig.\,\ref{Fig1}\,(d-g), as well as the different scaling methods used in the multifractal analysis \cite{Sgrignuoli_MF}.}
\label{Fig2}
\end{figure}

To demonstrate that the fluctuations of light scattered by the arrays exhibit strong multifractal properties we apply the statistical approach proposed by Chhabra and Jensen  \cite{Chhabra} and implemented in the FracLac ImageJ software package \cite{Schneider}. This approach is based on the box-counting method that 
divides the space embedding an object into a hyper-cubic grid of boxes of varying size $\epsilon$ (Supplemental Material for more details). We have computed three characteristic indicators of MF: (i) the generalized dimension $D(q)$, the mass exponent $\tau(q)$, and the multifractal spectrum $f(\alpha)$. The generalized dimension $D(q)$, first proposed as an alternative characterization of strange attractors of some dynamical systems, is defined as \cite{Grassberger}:
\begin{equation}\label{Dq}
D(q)=\frac{1}{q-1}\lim_{\epsilon\rightarrow0}\Biggl[\frac{\log \mu(q,\epsilon)}{\log \epsilon}\Biggr]
\end{equation}
where $q$ are the moments of a distribution and the partition function $\mu(q,\epsilon)$ is equal to
\begin{equation}\label{tau_definition}
\mu(q,\epsilon)=\sum_{k} P_k(\epsilon)^q\sim \epsilon^{-\tau(q)}
\end{equation} 
The probability $P_{k}(\epsilon)$ is evaluated as the integral of the measure over the $k^{th}$ box and it scales as $P_{k}(\epsilon)\sim \epsilon^{\alpha_k}$. The coefficients $\alpha_k$ are the Lipschitz-Holder exponents and quantify the strength of the singularity of a given positive measure \cite{Chhabra,Grassberger}. The mass exponent $\tau(q)$ describes the scaling of the partition function with respect to $\epsilon$ and it defines $D(q)$ through the relation $D(q)=\tau(q)/(q-1)$. Moreover, the Chhabra-Jensen method allows to define the one-parameter family $\hat{\mu}_k(q,\epsilon)$ through the relation  $\hat{\mu}_k(q,\epsilon)=P_k(\epsilon)^q/\sum_k P_k(\epsilon)^q$, such that the multifractal spectrum $f(\alpha)$ is obtained directly from the data by using the following expression:
\begin{equation}\label{f_ok}
f(\alpha)=\lim_{\epsilon\rightarrow0}\frac{\sum_k\hat{\mu}_i(q,\epsilon)\log[\hat{\mu}_i(q,l)]}{\log\epsilon}
\end{equation}
In particular, the numerator of Eqn.\,(\ref{f_ok}) is evaluated for each moments $q$ for decreasing box sizes. Then, $f(\alpha)$ is extrapolated from the slopes of $\sum_i\hat{\mu}_i(q,\epsilon)\log[\hat{\mu}_i(q,\epsilon)]$ as a function of $\log\epsilon$ \cite{Chhabra}.

Multifractal distributions are characterized by a smooth $D(q)$ function and a nonlinear dependence of $\tau$ from the moments $q$. These features are clearly visible in Figs.\,\ref{Fig2}\,(a-b). Moreover, the singularity spectra reported in Figs.\,\ref{Fig2}\,(c-d) exhibit a downward concavity with a large width $\Delta\alpha$, which is the hallmark of multifractality \cite{Chhabra,Nakayama}. The multifractal exponents extrapolated from the experimental data (lighter markers) follow well the ones estimated from the simulations (darker markers). Also note that multifractal spectra are global morphological properties of the diffraction spectra and are not influenced by the small local  intensity fluctuations due to sub-wavelength fabrication imperfections, as we show in more detail in the Supplemental Material.

Even though little is known about the singularity spectrum from an analytical point of view \cite{Vasquez}, the shape of $f(\alpha)$ encodes information on the PDF of the quantity under investigation \cite{Rodriguez}. In particular, it is known that if $f(\alpha)$ is parabolic (i.e., well reproduced by the model $f(\alpha)=a+b\alpha+c\alpha^2$ with three free parameters), then the PDF follows a log-normal statistic (i.e., a model characterized by two free parameters; the mean and standard deviation) \cite{Nakayama}. On the other hand, deviations of $f(\alpha)$ from a simple parabolic model reflect the non-Gaussian nature of its PDF \cite{Paladin,Vasquez,Rodriguez,Mandelbrot_3}. In this case, we speak of strong-MF \cite{Rodriguez}.

To better understand the multifractal properties of the diffraction intensity pattern of prime arrays, we have developed a non parabolic model characterized by the  polynomial function $f(\alpha)=p_1+p_2\alpha+p_3\alpha^2+p_4\alpha^3+p_5\alpha^4$. This model [continuous lines in Figs.\,\ref{Fig2}\,(c-d)] reproduces very well our data with a $R^2$-coefficient equal to 0.99 (see Supplemental Material for more details), demonstrating that photonic prime arrays exhibit strong-MF. Moreover, the nonparabolicity of the $f(\alpha)$ spectra indicates that the associated PDFs are non-Gaussian distributions characterized by heavy-tails.

\begin{figure}[t!]
\centering
\includegraphics[width=\columnwidth]{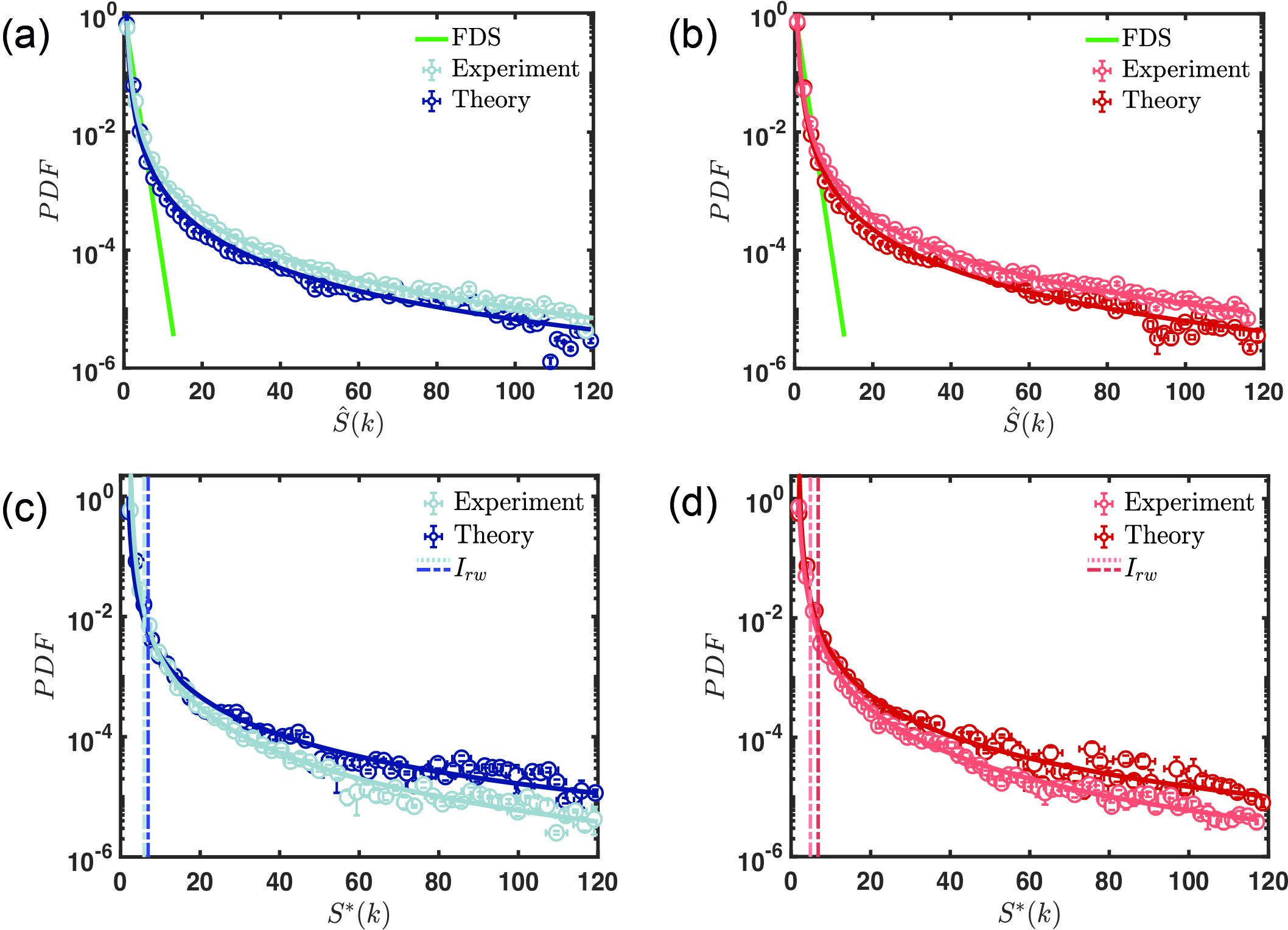}
\caption{Panels (a) and (b) display the PDFs of the fluctuations of the scattered radiation from the Eisenstein and Gaussian prime array. As a comparison, the negative exponential statistic of FDS, ensemble-averaged over 100 different uncorrelated arrangements of 2000 particles, is reported with a continuous-green line. PDFs of the most intense light intensity fluctuations produced by the Eisenstein (c) and Gaussian (d) prime array. The dashed lines indicate the threshold values $I_{rw}$. Markers represent the data: pastel-blue and pastel-red markers express the measured data, while navy-blue and dark-red markers refer to the calculated intensities. Continuous lines in panels (a-b) and (c-d) refer, respectively, to the best fits of both measured and simulated data obtained by using a least-square method based on eq.\,(\ref{Pareto}) and on eq.\,(\ref{GEV}).}
\label{Fig3}
\end{figure}

To demonstrate the link between strong-MF and non-Gaussian PDFs of the fluctuation of light intensity, we evaluate the histograms of the array structure factor normalized with respect to its averaged value $\hat{S}(k)=S(k)/\overline{S}(k)$. Figures\,\ref{Fig3}\,(a-b) show the results of this analysis extrapolated from the measured (lighter markers) and simulated (darker markers) data. In particular, these PDFs display remarkable heavy-tail features confirming the rogue wave character of the investigated far-field intensity profiles. The rare and extreme character of these distributions is particularly evident when compared against the negative exponential statistics $\exp[-\hat{S}(k)]$ (green continuous lines) that describes the fluctuations of waves scattered by uncorrelated random arrangements of particles, i.e., fully developed speckles (FDS) \cite{Goodman2}. 

To quantify the observed heavy-tail behavior, we use a least-square method based on the Pareto-type distribution defined as follows \cite{Pareto}:
\begin{equation}\label{Pareto}
P(x)=\frac{\beta}{\gamma\sigma}\left[1+\left(\frac{x-\mu}{\sigma}\right)^{1/2\gamma}\right]^{-(\beta+1)}\left(\frac{x-\mu}{\sigma}\right)^{-1+1/\gamma}
\end{equation}
where $x=\hat{S}(k)$, while $\mu$, $\sigma$, $\beta$, and $\gamma$ are four free parameters named, respectively, the location, the scale, the shape, and the inequality exponent. The distribution\,(\ref{Pareto}) reproduces the experimental and simulation data very well with a $R^2$-coefficient almost equal to 0.99.
These results, as discussed in more details in the Supplemental Material, are robust with respect to the number of scatterers (up to 10$^4$) and to the exposure time used to measure the diffraction intensities reported in Fig.\ref{Fig1}\,(d-e).The distributions of the Pareto family are $\alpha$-stable distributions often used in EVT to model hazards in Nature with long-range correlations, such as earthquakes, avalanches, rainfall distributions, the scaling laws of human travel, and financial crack markets \cite{Sharma,Brockmann,Longin,Leadbetter}. Our findings demonstrate their applications to the optical regime by linking the multifractal properties of the scattered radiation of photonic arrays with the heavy-tailed PDF\,(\ref{Pareto}). 

To demonstrate directly the optical rogue wave properties of the arrays, we compute histograms of the most intense fluctuations $S^*(k)$ and we evaluate the intensity threshold for rogue wave formation $I_{rw}$ by using the oceanographic definition $I_{rw} \geq 2I_{1/3}$. The quantity $I_{1/3}$ indicates the mean intensity of the highest third of events \cite{Dudley}. The results are reported in Figs.\,\ref{Fig3}\,(c-d), where the statistics extrapolated from the measured and simulated peak intensities are shown along with the intensity thresholds for rogue wave formation (dashed lines). The agreement between the statistics extracted from simulations (darker markers) and experiments (lighter markers) is remarkable. Moreover, these distributions are characterized by heavy-tails that exceed the threshold $I_{rw}$, demonstrating that the scattered radiation from the investigated arrays manifests the RW behavior. Concerning the probability of optical rogue wave events, Eisenstein and Gaussian prime arrays yield $3.6\pm 0.95\%$ and $4.1\pm1.3\%$ of events \footnote{The probability of rogue waves formation extracted from the simulations are compatible with the measured values, i.e., $(2.9\pm1.4)\%$  for the Eisenstein and $(3.4\pm1.4)\%$ for the Gaussian prime array. See also the Supplemental Material}. Thanks to the multifractality of their geometrical structures, the proposed devices support a larger probability for optical rogue wave generation as compared to what is observed in linear systems that impart a random phase on a coherent optical field \cite{Mathis}.

Finally, the continuous lines in Figs.\,\ref{Fig3}\,(c-d) show that the distributions of the most intense fluctuations of the scattered intensities are well reproduced, with a $R^2$-coefficient almost equal to 0.98, by the distribution $G(x)$ defined as \cite{Sharma}: 
\begin{multline}\label{GEV}
G(x)=\frac{1}{\rho}\left[1+\xi\left(\frac{x-\tau}{\rho}\right)\right]^{-(1+1/\xi)}\\
\exp\left\{-\left[1+\xi\left(\frac{x-\tau}{\rho}\right)\right]_+^{-1/\xi}\right\}
\end{multline}
where $+$ indicates the constraint $\left[1+\xi\left(\frac{x-\tau}{\rho}\right)\right]>0$, while $\xi$, $\rho$, and $\tau$ are the shape, the scale, and the location parameter, respectively. The probability density function (\ref{GEV}) is the compact form of three widely used distributions in EVT, named Gumbel (primarily adopted in corrosion engineering to guarantee the safety of buildings and in meteorological phenomena), Fr\'echet (mostly utilized to model market returns), and Weibull (originally developed in material science) \cite{Longin}. In fact, the extremal types theorem, which is the analog of the central limit theorem in EVT, states that the distribution of the rescaled maxima of extreme events belongs to one of these three families \cite{Coles,Leadbetter}. Collectively, these three distributions are named extreme value distributions and can be parametrized into the single family distribution (\ref{GEV}) known as the generalized extreme value distribution \cite{Coles,Leadbetter}. The shape parameter $\xi$  discriminates the distribution class. $G(x)$ belongs to the Gumbel, to the Weibull, or to the Fr\'echet class depending whether $\xi$ is equal to zero, lower than zero, or larger than zero, respectively. Moreover, a direct link exists between the distribution of extreme events $P(x)$ and the distribution of their most intense values $G(x)$ \cite{Coles,Leadbetter,Longin,Sharma}. Such link is particularly used in finance to predict the price movements characterized by independent and stationary increments, i.e., L\'evy processes \cite{Longin}. In particular, if $P(x)$ corresponds to the normal or exponential distribution, the distribution of the maxima values follows a Gumbel function. In cases where $P(x)$ is characterized by a finite support (i.e., upper endpoint such as the uniform and the beta distribution), $G(x)$ approaches to the Weibull distribution. Lastly, the $G(x)$ falls into the Fr\'echet class when $P(x)$ exhibits power-law (i.e., Pareto-type) tails \cite{Leadbetter,Longin,Sharma}. The data of Figs.\,\ref{Fig3}\,(c-d) are well-described for $\xi$ values always larger than zero (see Supplemental Material) in agreement with the power-law features of the PDFs of panels\,(a-b). Interestingly, extreme value Fr\'echet distributions were observed in the statistics of natural hazards with long-term memory characteristics \cite{Sharma,Longin} and in the scaling behavior of focused waves in random media modeled as a Gaussian random field \cite{Metzger}. Our findings generalize these results to linear optical systems with strong multifractal behavior, introducing a novel approach to achieve high intensity optical hot spots. 


In conclusion, using the approach of extreme value theory we demonstrated the formation of optical RWs with enhanced probability in multifractal arrays of dielectric nanoparticles based on the distributions of the prime elements in complex quadratic fields (prime arrays). Our findings link directly the strong multifractality of photonic prime arrays with intense RW phenomena in the linear optical regime. In particular, we developed a non-parabolic multifractal model that well-reproduces the experimentally measured multifractal spectra for the scattered radiation. The non-Gaussian PDFs of light intensity fluctuations associated to strong-multifractality 
have been characterized using heavy-tailed Pareto-type and the Fr\'echet distribution for the most intense values, in excellent agreement with the experimental results. Notably, multifractality characterizes also oceanic rogue waves \cite{Hadjihosseini}, suggesting a novel interpretation in the long-debated analogy between oceanographic and optical rogue waves. Our results unveil the key role of long-range structural correlations and strong-multifractality for the engineering of planar diffraction arrays that produce high-intensity RWs for applications to enhanced sensing and lithography.

\begin{acknowledgments}
L.D.N. acknowledges the partial support from National Science Foundation (ECCS-2015700), the partial support from National Science Foundation (DMR-1709704), and the partial support from the Army Research Laboratory (ARL; Cooperative Agreement Number W911NF-12-2-0023) for the development of theoretical modeling.
\end{acknowledgments}
\section*{Author contributions}
F.S. performed numerical calculations, data analysis, and organized the results with the help of Y. C. S.G. performed the experiments together with F.S. W.A.B. fabricated the samples. L.D.N. conceived and led the work. F.S. and L.D.N. wrote the manuscript.
%
\section{Supplementary Information}
\subsection{Sample fabrication}
\begin{figure*}[h!]
\centering
\includegraphics[width=\textwidth]{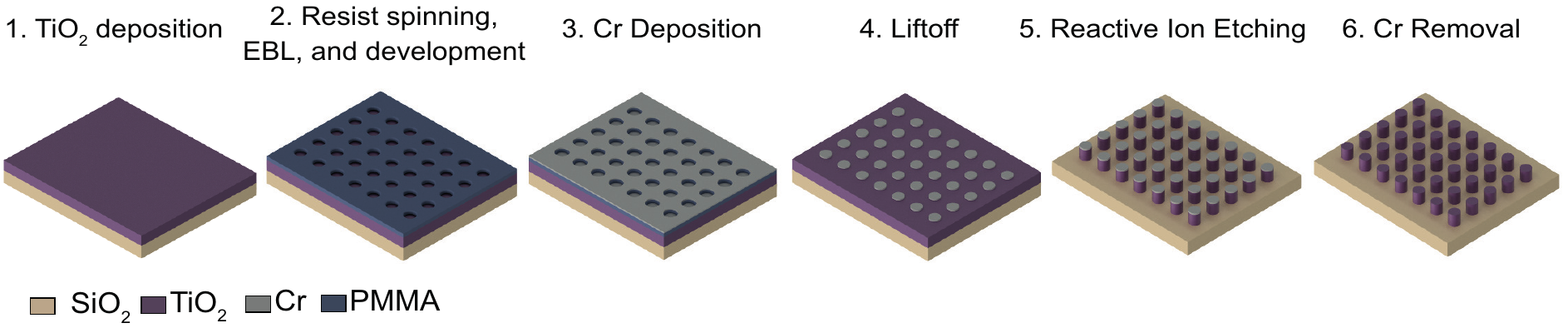}
\caption{Process flow utilized to fabricate the Eisenstein and Gaussian prime arrays.}
\label{FigS0}
\end{figure*}

Figure\,\ref{FigS0} shows the process flow utilized to fabricate the investigated photonic prime arrays. As a first step, TiO$_2$ thin films were deposit by reactive direct current (DC) magnetron sputtering (MSP) on quartz (SiO$_2$) substrates with a Denton Discovery D8 Sputtering System. As a second step, a 100\,nm thick polymethyl methacrylate (PMMA) resist layer was spun and baked before sputtering a thin conducting layer ($\sim$6\,nm) of Au to compensate for the electrically insulating SiO$_2$ substrate. The resist was exposed at 30 keV by an SEM (Zeiss Supra 40) integrated with Nanometer Pattern Generation System direct-write software. Electron beam lithography was then used to drill nanoholes following the spatial distribution of Eisenstein and Gaussian prime elements. The obtained samples were developed in isopropyl alcohol to methyl isobutyl ketone (IPA:MIBK) (3:1) solution for 70\,sec followed by rinsing with IPA for 20\,sec. As a third step, we deposited a 20\,nm thick layer of Cr on top of the developed resist by using electron beam evaporation (CHA Industries Solution System). After this deposition, we removed unwanted Cr with a three minutes lift-off in acetone, and the nanoparticle patterns were transferred from the Cr mask to the TiO$_2$ thin film by reactive ion etching (RIE, Plasma-Therm, model 790) using Ar and SF$_6$ gases. As final step, the Cr mask was removed by wet etching in Transene 1020 \cite{Zheng}. More details can be found in Ref.\,\cite{Sgrignuoli_MF}.

To characterize the quality of the fabricated samples, we have estimated the surface roughness and the experimental fluctuations in the position and size of the nanocylinders. The fabricated arrays are fabricated from a 250nm TiO$_2$ thin film with an intrinsic 0.5nm rms surface roughness when measured by optical profilometry on a (100) silicon substrate. The fabrication error on the positions of the particles  depends on the error of the coils that direct the ELB rastering unit. Our ELB system is driven by 16-bit defection coils that yield a 1.2\,nm in-plane displacement error when writing a 55\,$\mu$m $\times$ 55\,$\mu$m square field of view. As discussed in Ref.\,\cite{Sgrignuoli_MF}, the fabricated TiO$_2$ nanocylinders with a 110\,nm target radius have a size dispersion profile characterized by a Gaussian shape. This feature is associated with the random fluctuations of the electron beam system and the diffusion of the resist during the development. We have estimated a 9\% fluctuation error in the radius of the fabricated nanocylinders with respect to the target value. However, these unavoidable fabrication imperfections are deeply sub-wavelength and do not significantly perturb the main results of our work. This is demonstrated in Fig.\,\ref{FigSEM}\,(a-b), where we show that the SEM images of the fabricated samples match very well the position of the designed nanocylinders (red circle markers). The observed systematic shift at the edges of Fig. R2 (a-b) is an artifact of the electronic imaging process. This fact is related to the time-dependent buildup of electronic charge on the sample that induces a systematic global repulsion of the electron beam \cite{Maraghechi}.
\subsection{Box counting method}
To characterize the size-scaling features of the scattered radiation from prime arrays, we employ the box-counting method, which can be briefly described as follows. This approach subdivides the object under investigation into non-overlapping hyper-cubic boxes of different sizes $\epsilon$. Then, the minimum number of boxes $N(\epsilon)$ needed to cover the system for each considered $\epsilon$ is determined. While homogeneous fractals are characterized by a global scale-invariant symmetry described by the power-law scaling $N(\epsilon)\sim \epsilon^{-D_f}$, multifractals are described by a hierarchy of fractal dimensions $D_f$ through the continuous function $f(\alpha)$, named multifractal spectrum. For multifractals, the minimum number of boxes $N(\epsilon)$  scales as $\epsilon^{-f(\alpha)}$ \cite{Nakayama}.

\begin{figure*}[h!]
\centering
\includegraphics[width=12cm]{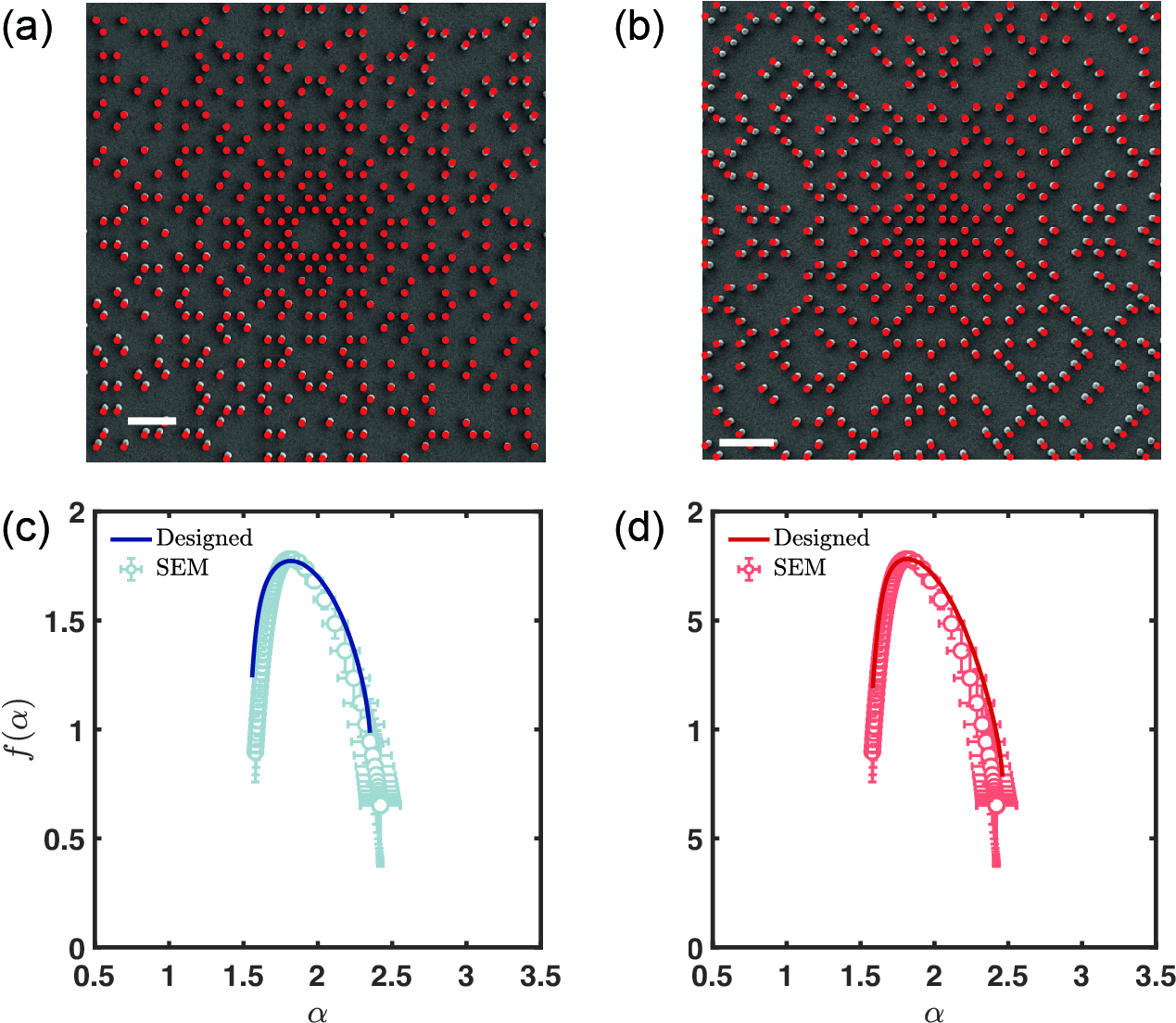}
\caption{Panel (a) and (b) show the SEM images of the fabricated Eisenstein and Gaussian prime arrays, respectively, superimposed to the corresponding designed point patterns (red circle markers). The white bars correspond to 2 $\mu$m. Panel (c) and (d) display the $f(\alpha)$ spectra as a function of the size-scaling exponent $\alpha$ extrapolated from the respective sample at the top of each column.}
\label{FigSEM}
\end{figure*}

Multifractal spectra and parameters characterize global morphological properties of complex systems and are not influenced by sub-wavelength fabrication imperfections. Figure\,\ref{FigSEM}\,(c-d) shows that the $f(\alpha)$ exponents extrapolated from the designed arrays are well within the error bars of the spectra derived from the SEM images.
\subsection{Parabolic approximation of $f(\alpha)$ and strong multifractality}
\begin{figure*}[b!]
\centering
\includegraphics[width=14cm]{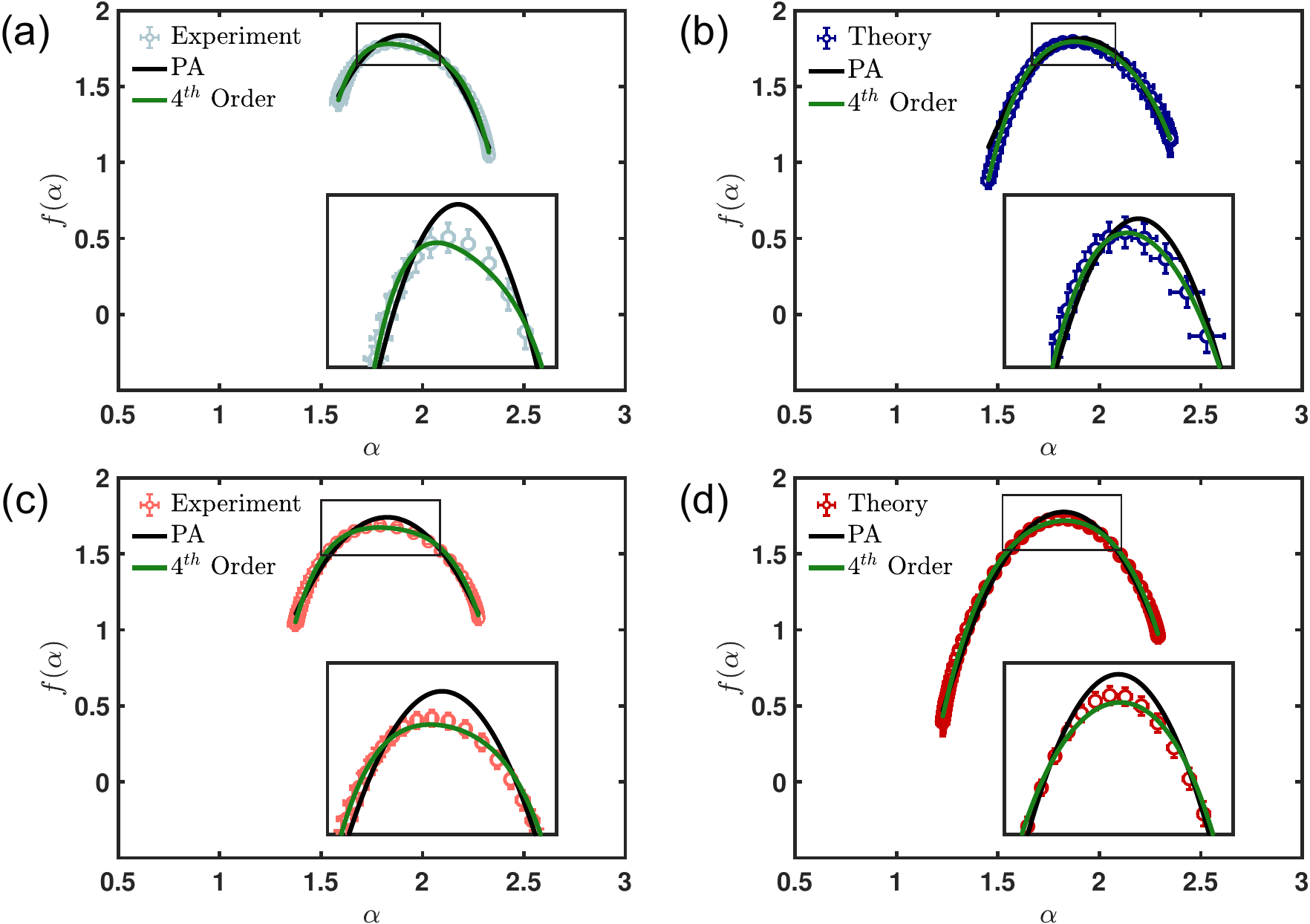}
\caption{Panels (a-b) and (c-d) show the multifractal spectra $f(\alpha)$, as a function of the local scaling exponent $\alpha$, extrapolated from the radiation scattered from Eisenstein and Gaussian prime arrays. Panels (a-c) and (b-d) refer to the measured and simulated data. Markers represent the data. The black and green continuous lines refer to the best fits obtained employing a least-square procedure based on the parabolic approximation and the model (S2), respectively. Error bars take into account the different threshold percentages (between 55\% and 75\% of the maximum intensity value) used to binarize the diffraction patterns of Fig.\,1\,(d-g) of the main manuscript, as well as the different scaling methods used in the multifractal analysis \cite{Sgrignuoli_MF}.}
\label{FigS1}
\end{figure*}

The shape of the multifractal spectrum $f(\alpha)$ contains information on the statistic of the quantity $\psi$ under investigation \cite{Rodriguez,Vasquez}; in the following identified with the diffraction intensity. The parabolic approximation (PA) for $f(\alpha)$, defined by the expression \cite{Nakayama}
\begin{equation}\label{PA_eq}
f(\alpha)_{PA}=a+b\alpha+c\alpha^2~~\mbox{,}
\end{equation}
implies that the probability distribution function (PDF) of $\psi$ follows a log-normal statistic. In eq.\,(\ref{PA_eq}) $\alpha$ are the local scaling exponents, while $a$, $b$, and $c$ are fitting parameters. We use a least-square method to test the deviation from the parabolic approximation of the multifractal spectra discussed in the main manuscript and also reported in Fig.\ref{FigS1}. In Fig.\ref{FigS1} the lighter and darker markers indicate, respectively, the $f(\alpha)$ extrapolated from the measured and simulated data. The results of the fit analysis are summarized in Table\,\ref{PA} and displayed in Fig.\ref{FigS1} with continuous black lines. These findings demonstrate the nonparabolicity of the investigated spectra. The model \,(\ref{PA_eq}) does not reproduce the behavior of the investigated multifractal spectra especially nearby the peak values $f(\alpha_0)=D_f$, as shown in the insets of Fig.\ref{FigS1}.
\begin{table}[t!]
\caption{The $R^2$ coefficient and the fitting parameters $a$, $b$, and $c$ of the parabolic model (\ref{PA_eq}) are displayed for both the measured and simulated multifractal spectra $f(\alpha)$ of the Eisenstein (EP) and Gaussian (GP) prime array.}
\centering
 \begin{tabular}{c || c c | c c |} 
 \hline \hline
            & EP$_{experiment}$ & EP$_{theory}$  & GP$_{experiment}$ &GP$_{theory}$ \\
 \hline \hline \\ 
 R$^2$ & 0.976                       & 0.982                & 0.975                        & 0.981 \\ [1ex] 
 a         & -12.79                      & -10.7                 & -8.71                        & -10.5 \\ [1ex] 
 b         & 15.4                         & 13.09                &  11.45                       & 13.45 \\[1ex]
 c         & -4.05                        & -3.43                 &   -3.14                      & -3.69 \\[1ex] 
 \hline
 \end{tabular}
 \label{PA}
\end{table}
\begin{table}[b!]
\caption{The $R^2$ coefficient and the fitting parameters $p_j$ ($j=1,\dots,5$) of the fourth-order polynomial function (\ref{4order}) are reported for both the measured and simulated multifractal spectra $f(\alpha)$ of the Eisenstein (EP) and Gaussian (GP) prime array.}
\centering
 \begin{tabular}{c || c c | c c |} 
 \hline \hline
            & EP$_{experiment}$ & EP$_{theory}$  & GP$_{experiment}$ &GP$_{theory}$ \\
 \hline \hline \\ 
 R$^2$ & 0.998                      & 0.999                 & 0.997                       & 0.998 \\ [1ex] 
 p$_1$ & -280.4                     & -107.4                & -105.3                      & -56.17 \\ [1ex] 
 p$_2$ & 576.1                      & 212.7                 & 228.5                        & 111.4 \\[1ex]
 p$_3$ & -441.7                     & -156.7                & -183.8                      & -88.38 \\[1ex] 
 p$_4$& 150.8                       & 51.89                 & 66.06                        & 31.97 \\[1ex] 
 p$_5$ & -19.34                     & -6.541                & -8.958                      & -4.453 \\ [1ex] 
 \hline
 \end{tabular}
 \label{Table_4Ord}
\end{table}

To better understand the multifractal properties of the scattered radiation from prime arrays, we have developed a model characterized by the fourth-order polynomial function: 
\begin{equation}\label{4order}
f(\alpha)=p_1+p_2\alpha+p_3\alpha^2+p_4\alpha^3+p_5\alpha^4
\end{equation}
where $p_j$ ($j=1\cdots5$) are free parameters. The model (\ref{4order}) describes very-well the investigated $f(\alpha)$ spectra with an $R^2$-coefficient very close to one, as summarized in Table\,\ref{Table_4Ord} and highlighted in Fig.\,\ref{FigS1} with the green-continuous lines. These findings demonstrate the strong multifractal features and the non-Gaussian nature of the investigated spectra. 

\subsection{Analysis of pair distribution functions}
\begin{figure*}[b!]
\centering
\includegraphics[width=14.5cm]{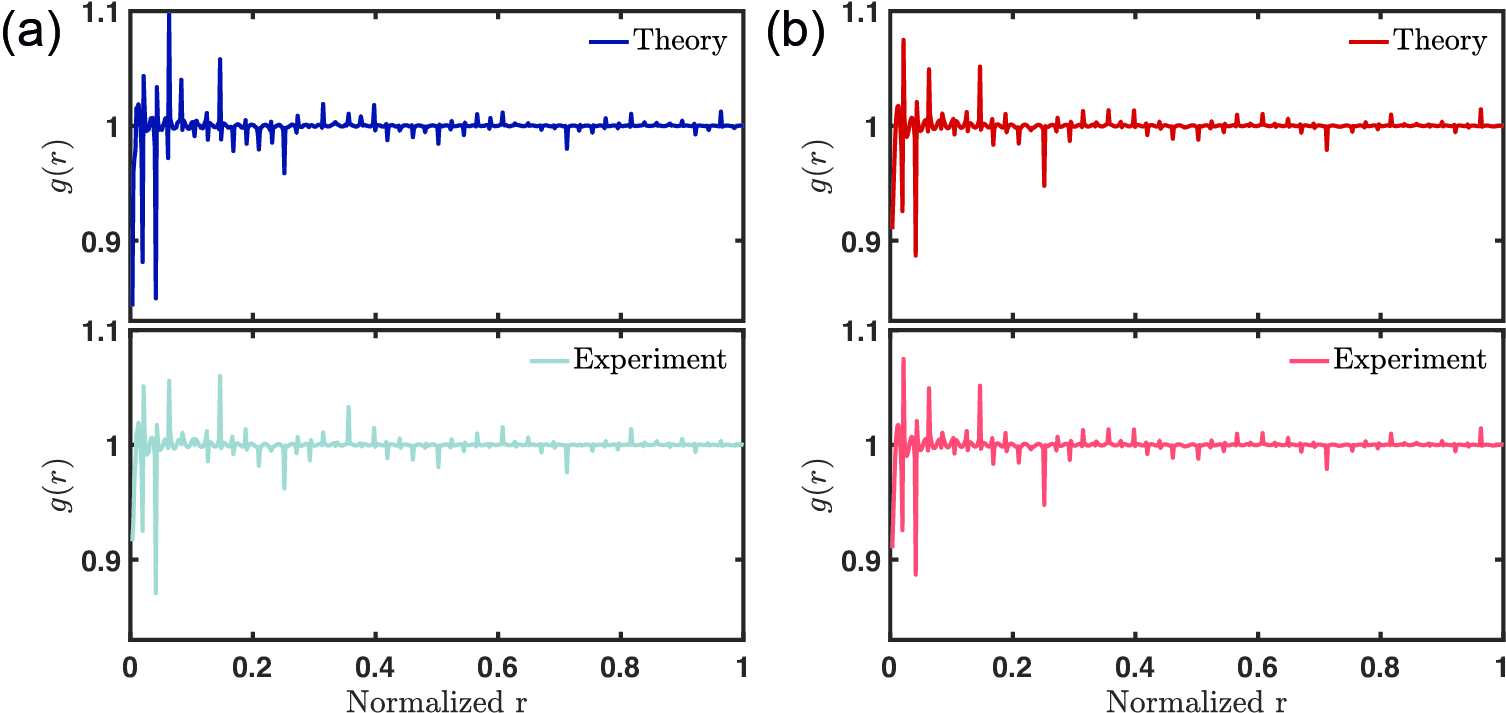}
\caption{Panel (a) and (b) show the Eisenstein and Gaussian prime arrays\textsc{\char13} pair distribution function, respectively. Top panels are evaluated from the simulated structure factors, while the bottom panels from the measured diffraction intensity profiles.}
\label{Fig_g2}
\end{figure*}

To provide a quantitative metric and establish a direct correspondence between the measured and simulated far-field diffraction patterns, we have evaluated the pair distribution function $g(r)$ from the diffraction spectra. This function establishes a quantitatively assess of the quality of the diffraction patterns and it is used in crystallography to gain quantitative insight into the atomic-scale properties of materials that do not exhibit long-range periodicity, such as covalent glasses and amorphous semiconductors \cite{Cliffe,Billinge,Janot}. The pair distribution function is deduced, under the assumptions of homogeneity and isotropy, from the structure factor $S(k)$ through the formula \cite{Janot}:
\begin{equation}\label{g2Pair}
g(r)=1+\frac{1}{(2\pi)^3n}\int_0^\infty [S(k)-1]\frac{\sin kr}{kr} 4\pi k^2dk
\end{equation}
where $n$ is the particle density. To apply eq.\,(\ref{g2Pair}) in our case, we have azimuthally averaged the diffraction intensity reported in Fig.1 of the main manuscript. Figures \ref{Fig_g2} (a-b) compare the pair distribution function obtained from the simulated (top panels) and the measured (bottom panels) structure factors of Eisenstein and Gaussian prime arrays, respectively. The R$^2$-coefficient between the measured and simulated $g(r)$ curves is 82\% for the Eisenstein and 95\% for Gaussian. This quantitative metric establishes an additional comparison (see also Fig.2 and Fig.3 and the relative discussions in the main manuscript) between the measured and computed diffraction spectra and demonstrates the very high quality of the fabricated samples.

\subsection{Heavy-tail distributions: fitting results and trends stability}
\begin{table}[t!]
\caption{The $R^2$ coefficient and the fitting parameters $\mu$, $\sigma$, $\gamma$, and $\beta$ (named, respectively, the location, the scale, the shape, and the inequality exponents) of the distribution (\ref{Pareto}) are displayed for both the measured and simulated light intensity fluctuations generated by the Eisenstein (EP) and Gaussian (GP) prime array.}
\centering
 \begin{tabular}{c || c c | c c |} 
 \hline \hline
                   & EP$_{experiment}$ & EP$_{theory}$  & GP$_{experiment}$ &GP$_{theory}$ \\
 \hline \hline \\ 
 R$^2$        & 0.99                        & 0.98                   & 0.99                         & 0.98 \\ [1ex] 
 $\mu$        & 0.31                        & 0.30                   & 0.24                         & 0.29 \\ [1ex] 
 $\sigma$   & 0.07                        & 0.05                   & 0.05                         & 0.05 \\[1ex]
 $\gamma$ & 0.39                        & 0.39                  & 0.39                         & 0.39 \\[1ex] 
 $\beta$      & 1.91                        & 1.91                  & 1.85                          & 1.91 \\[1ex] 
 \hline
 \end{tabular}
 \label{Table_Pareto}
\end{table}

As discussed in the main manuscript, we discovered that the PDFs of the fluctuations of the scattered light from prime arrays are well reproduced by the Pareto-type distribution: 
\begin{equation}\label{Pareto}
P(x)=\frac{\beta}{\gamma\sigma}\left[1+\left(\frac{x-\mu}{\sigma}\right)^{1/2\gamma}\right]^{-(\beta+1)}\left(\frac{x-\mu}{\sigma}\right)^{-1+1/\gamma}
\end{equation}
The distributions of the Pareto-family are often used in extreme value theory (EVT) to model extreme events in Nature and society characterized by multifractal features \cite{Sharma,Leadbetter,Longin}. Table \ref{Table_Pareto} reports the results of the least-square fit method based on the model (\ref{Pareto}) applied to the distributions of both measured and simulated data. The values of the R$^2$-coefficient nearby one demonstrate the non-Gaussian nature of the investigated PDFs. Moreover, these findings prove that the scattered radiation of the investigated arrays encodes rare and extreme rogue wave events. 

The extremal types theorem \cite{Coles,Leadbetter}, which is the analog of the central limit theorem in EVT, establishes that if the PDF of rare and extreme events is described by a Pareto-type distribution, then the distribution of their most intense values has to follow a Fr\'echet statistic \cite{Sharma,Leadbetter,Longin,Coles}. To prove this fact, we have used the generalized extreme value distribution \cite{Sharma}
\begin{table}[b!]
\caption{The $R^2$ coefficient and the fitting parameters $\tau$, $\rho$, and $\xi$ (named, respectively, the location, the scale, and the shape exponents) of the model (\ref{GEV}) are summarized for both the measured and simulated distribution of the most intense fluctuations of the scattered intensity characterizing the Eisenstein (EP) and Gaussian (GP) prime array.}
\centering
 \begin{tabular}{c || c c | c c |} 
 \hline \hline
                   & EP$_{experiment}$ & EP$_{theory}$  & GP$_{experiment}$ &GP$_{theory}$ \\
 \hline \hline \\ 
 R$^2$        & 0.98                        & 0.97                  & 0.99                         & 0.97 \\ [1ex] 
 $\tau$        & 2.66                        & 1.85                  & 1.79                         & 2.16 \\ [1ex] 
 $\rho$       & 0.21                        & 0.15                   & 0.23                         & 0.20 \\[1ex]
 $\xi$         & 0.76                        & 1                         & 0.75                         & 0.92 \\[1ex] 
 \hline
 \end{tabular}
 \label{GEVTable}
\end{table}
\begin{multline}\label{GEV}
G(x)=\frac{1}{\rho}\left[1+\xi\left(\frac{x-\tau}{\rho}\right)\right]^{-(1+1/\xi)}\\
\exp\left\{-\left[1+\xi\left(\frac{x-\tau}{\rho}\right)\right]_+^{-1/\xi}\right\}
\end{multline}
to reproduce the histograms of the most intense scattered radiation events, detected by using an 8-connected neighborhood local maximum search applied to the Fig.\,1\,(d-g) of the main manuscript \cite{Mathis}. Table \ref{GEVTable} reports the results of the least-square fit method based on the model (\ref{GEV}) applied to both the measured and simulated data. The $R^2$-coefficient values are nearby the unity, and the fact that the $\xi$ values are always larger than zero demonstrates that the investigated distributions belong indeed to the Fr\'echet class, as discussed in more detail in the main manuscript. 

\begin{figure*}[b!]
\centering
\includegraphics[width=14cm]{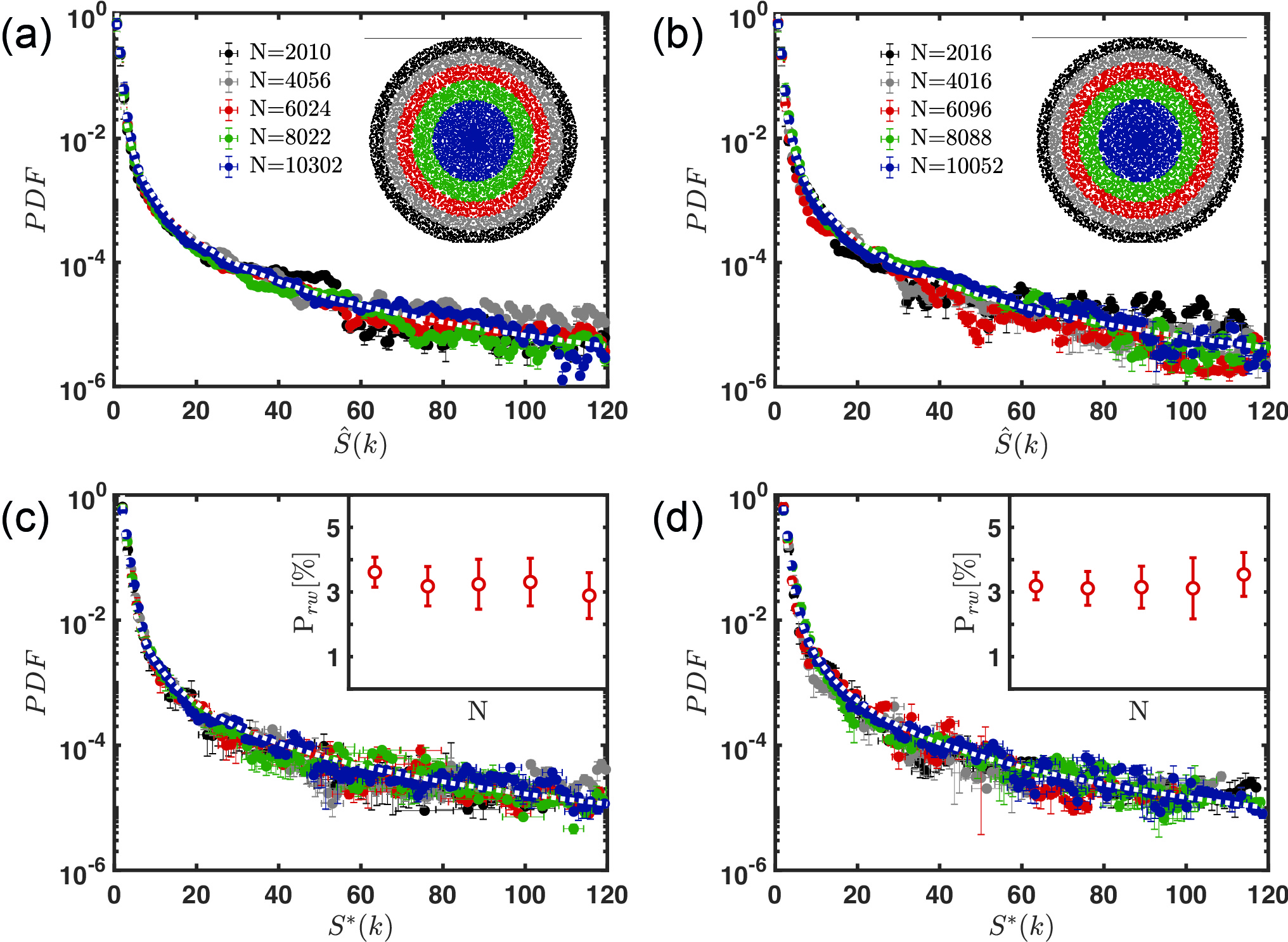}
\caption{Panels (a) and (b) display the scaling of the PDFs of the fluctuations of the scattered radiation from Eisenstein (a) and Gaussian (b) arrays characterized by different prime elements, as specified in the legends. Specifically, the black, gray, red, green, and blue markers refer to an array composed of almost 2000, 4000, 6000, 8000, and 10000 scatterers. Panels (c) and (d) report the scaling of the distributions of the most intense light intensity fluctuations produced by the Eisenstein and Gaussian configuration. The dotted white lines in panels (a-b) and (c-d) are the best fits obtained by employ a least-square method based on the distributions (\ref{Pareto}) and (\ref{GEV}), respectively, applied to the most extensive array configuration. The insets of panels (c-d) display the probability of rogue wave events $P_{rw}$ with respect to the number of prime elements $N$. Error bars are the standard deviations with respect to the different bin values used to perform the histograms. }
\label{FigS2}
\end{figure*}

Figure \,\ref{FigS2}\, shows the PDFs of the fluctuations of the simulated diffracted intensity (a-b), as well as  the distributions of their most intense events (c-d), by increasing the number of Eisenstein (a-c) and Gaussian (b-d) prime elements. These heavy-tail distributions remain approximately the same by increasing the number of scatterers from 2000 up to 10000. Moreover, they are well-reproduced by the model (\ref{Pareto}) [dotted-white lines in Figs.\,\ref{FigS2}(a-b)] and by the distribution (\ref{GEV}) [dotted-white lines in Figs.\,\ref{FigS2}(c-d)] applied to the most extensive size array configuration. This specific configuration is the one discussed in Fig.\,3 of the main manuscript. Error bars are the standard deviations with respect to the different bin values used for each array size to perform the histograms. The probability of optical rogue formation $P_{rw}$, evaluated by using the oceanographic definition $I_{rw} \geq 2I_{1/3}$ \cite{Dudley,Dudley_2}, is almost independent of the number of prime elements, as reported in the insets of Fig.\,\ref{FigS2}.

\begin{figure*}[b!]
\centering
\includegraphics[width=14cm]{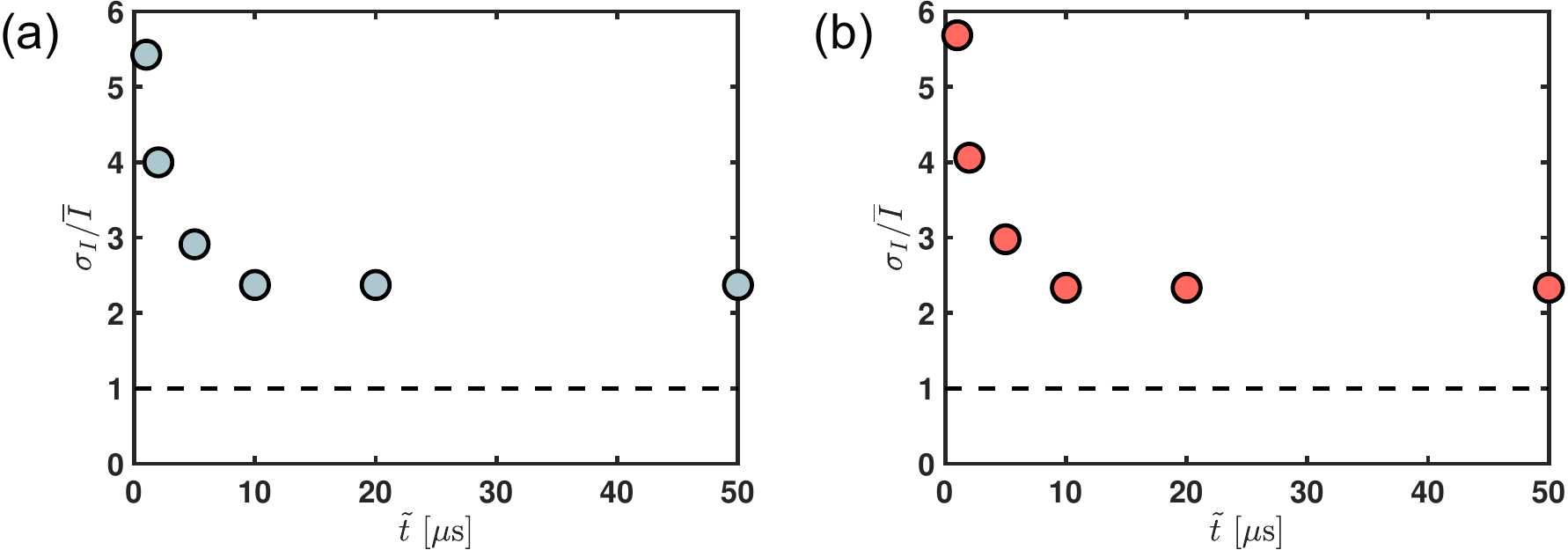}
\caption{Panels (a) and (b) show the contrast factor C as a function of the exposure time $\tilde{t}$, extrapolated from the measured intensities profile of the Eisenstein and Gaussian (d) prime array. The black dashed lines indicate the contrast factor of a transmitted intensity pattern that obeys the Gaussian statistic. }
\label{FigS3}
\end{figure*}

To investigate the effects of the exposure time $\tilde{t}$ used during the measurements, we have evaluated the contrast factor $C$ for different $\tilde{t}$. This parameter is defined as the ratio of standard deviation $\sigma_I$ of intensity to the mean intensity $\overline{I}$ \cite{Goodman2}:
\begin{equation}
C=\frac{\sigma_I}{\overline{I}}
\end{equation}
Figures\,\ref{FigS3}\,(a-b) display the results of this analysis for the Eisenstein and Gaussian configuration, respectively. Transmitted intensity patterns that obey the Gaussian statistic have a contrast factor equal to one, implying that the fluctuations of the intensity are comparable to the mean value. On the other hand, a $C$ factor above unity indicates the presence of sharp focusing effects \cite{Safari}. For this reason, the contrast factor, or its square value known as the scintillation index, are used to characterize the sharpness of the caustics that exhibit rogue-type behavior in both linear and nonlinear regime \cite{Safari}. The trends of the $C$ factor, shown in Figs.\,\ref{FigS3}\,(a-b), are always larger than unity, demonstrating the non-Gaussian nature of the measured scattered radiation. Moreover, the contrast factor $C$ saturates after 10\,$\mu s$ around a value larger than 2, showing that the exposure time does not influence the rogue waves generation.
\end{document}